\documentclass[graybox]{svmult}

\usepackage{type1cm}        
\usepackage{makeidx}         
\usepackage{graphicx}        
\usepackage{multicol}        
\usepackage[bottom]{footmisc}

\makeindex             

\begin{document}

\title*{Expansion by regions: an overview}
\author{Vladimir A. Smirnov}
\institute{Vladimir A.\ Smirnov \at Skobeltsyn Institute of Nuclear Physics of Moscow State 
University, 119992 Moscow, Russia, \email{smirnov@theory.sinp.msu.ru}. 
The paper is based on the talk at
``Antidifferentiation and the Calculation of Feynman Amplitudes`` (4-9 October 2020, Zeuthen, Germany)
and published in the book with this title (Springer 2021, doi:10.1007/978-3-030-80219-6).
}
%
%
\maketitle


\abstract{A short review of expansion by regions is presented. It is a well-known strategy to 
obtain an expansion of a given
multiloop Feynman integral in a given limit where some kinematic invariants and/or masses
have certain scaling measured in powers of a given small parameter. Prescriptions of this strategy
are formulated in a simple geometrical language and are illustrated through simple examples.}

\section{Historiographical notes}
\label{sec:1}

Expansion by regions is a universal strategy to obtain an expansion of a given
Feynman integral in a given limit, where kinematic invariants and/or masses essentially differ in scale.
For simplicity, let us consider a Feynman integral $G_{\Gamma}(q^2,m^2)$ depending on two scales, for example, $q^2$ and $m^2$, and let the limit be $t=-m^2/q^2\to 0$. 
Experience tells us that the expansion at $t\to 0$ has the form
\begin{equation}
 G_{\Gamma}(t,\varepsilon) \sim (-q^2)^{\omega}
\sum_{n=n_0}^{\infty}\sum_{k=0}^{2h} c_{n,k}(\varepsilon)\, t^n\,\log^k t \;, 
\label{exp1}
\end{equation}
where 
$\omega=4h-2\sum a_i$ is the degree of divergence, with $a_l$ powers of the propagators, 
$h$ is the number of loops and $\varepsilon=(4-d)/2$ is the parameter of dimensional regularization. 
The expansion is often called asymptotic, in the sense that the remainder of expansion has the order $o(t^N)$ after
keeping terms up to $t^N$ . However, every power series at a power of logarithm in expansions in various limits of momenta and masses has a non-zero radius of convergence which is determined usually by 
the nearest threshold.

There can be different reasons to consider some limit and the corresponding expansion. Typically,
different scaling of kinematic invariants and/or masses involved is dictated by a phenomenological situation.
Moreover, experience obtained when expanding Feynman integrals in some limit can show a way to construct
the corresponding effective theory. At the level of individual Feynman integrals, expanding a complicated Feynman integral in some limit can approximately substitute the analytic evaluation of the integral.
 
One can use various techniques in order to obtain an expansion of a given Feynman integral in some limit: one can start with a parametric representation, or apply the method of Mellin--Barnes representation, or obtain an expansion within the method of differential equations. However, the {\em general} strategy of expansion by regions provides the possibility to write down a result for the expansion immediately once relevant regions are known. Such a result looks similar to
(\ref{exp1}) but now exponents of the expansion parameter depending linearly on $\varepsilon$ are not yet expanded in 
$\varepsilon$,
\begin{equation}
 G_{\Gamma}(t,\varepsilon) \sim (-q^2)^{\omega}
\sum_{n=n_0}^{\infty}\sum_{k=0}^{h}\sum_{j=0}^{h} c'_{n,j,k}(\varepsilon) t^{n-j\varepsilon}  \log^k t 
\label{exp2}
\end{equation}
and the coefficients in the expansion can be represented in terms of integrals over loop momenta
or over Feynman parameters. These integrals on the right-hand side of the expansion are constructed
according to certain rules starting from the Feynman integral or a parametric integral for the initial
Feynman integrals $G_{\Gamma}$. This means that expansion by regions reduces the problem to the
evaluation of integrals present in (\ref{exp2}).

Logarithms in (\ref{exp2}) within dimensional regularization do not appear in limits typical of Euclidean space such as
the off-shell large momentum limit and the large mass limit. Rather, they are typical for limits typical of
Minkowski space such as the Regge limit and various versions of the Sudakov limit. In fact, one can avoid such logarithms by introducing an auxiliary analytic regularization which can be introduced as additional complex numbers in the exponents of the propagators. One can say that, after this, the various scales in the problem become separated so that the expansion becomes only in powers of the expansion parameter. After turning off this regularization, spurious poles in the auxiliary analytic parameters cancel giving rise to the logarithms, and this happens to be an important consistency check. A lot of examples illustrating this phenomenon can be found, e.g., in \cite{Smirnov:2002pj}.
We will come back to this point in Section~2 when discussing the geometrical formulation of expansion by regions.

According to the first formulation of  expansion by regions \cite{Beneke:1997zp} one analyzes various regions in a given integral over loop momenta 
and, in every region, expands the integrand 
in parameters which are there small. Then the integration in the integral with so expanded 
propagators is extended to the whole domain of the loop momenta and, finally, one obtains
an expansion of the given integral as the corresponding sum over the regions. Although these recipes were formulated in a suspicious mathematical language, expansion by regions was successfully applied in numerous calculations.

\begin{figure}[ht]
\sidecaption[ht]
\includegraphics[scale=1.]{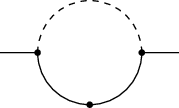}
\caption{A one-loop graph.}
\label{fig:1}        
\end{figure}
A very simple example is given by the Feynman integral corresponding to the graph depicted in Fig.\ref{fig:1},
\begin{equation}
 G(q^2,m^2;d) =
\int \frac{{\rm d}^d k}{(k^{2} -m^{2})^2 (q-k)^{2}} 
\label{ex1}
\end{equation}
in the limit $m^2/q^2 \to 0$. 

The relevant regions are the region of small loop momenta, $k\sim m$, and 
the region of large loop momenta, $k\sim q$. According to the above prescriptions,
in the first region, the first propagator is unexpanded and
the second propagator is expanded in a Taylor series in $k$. In the second region, 
the first propagator is expanded in a Taylor series in $m$ the second propagator is unexpanded.
The leading terms of expansion give
\begin{equation}
 G(q^2,m^2;d) \sim \int \frac{{\rm d}^d k}{(k^{2})^2 (q-k)^{2}}
+\frac{1}{q^2} \int \frac{{\rm d}^d k}{(k^{2} -m^{2})^2}
   + \ldots 
   \label{f2}
\end{equation}
The integrals involved can be evaluated by Feynman parameters, with the following result
\begin{equation}
G(q^2,m^2;d) \sim
i \pi^{d/2}\left( \frac{\Gamma(1-\varepsilon)^2\Gamma(\varepsilon)}{\Gamma(1-2\varepsilon)(-q^2)^{1+\varepsilon}}
+\frac{\Gamma(\varepsilon)}{q^2(m^2)^{\varepsilon}} + \ldots\right) 
\label{f3}
\end{equation}

Although the initial Feynman integral is finite at $d=4$, there are simple poles above:
an infrared pole in the first term and an ultraviolet term in the second term. They are successfully canceled,
with the following result
\begin{equation}
 i \pi^{d/2} \left(\log\left(\frac{-q^2}{m^2} \right) + \ldots \right)\;.
\end{equation}

Such an interplay of various divergences is a typical feature of expansions in momenta and masses. Only in rare situations, such as an expansion in the small momentum limit of a Feynman integral without massless threshold in the corresponding channel, there is no such phenomenon. Let me also point out that the first term in (\ref{f2}) is convergent
at $\mbox{Re}(\varepsilon)<0$ while the second term in (\ref{f2}) is convergent
at $\mbox{Re}(\varepsilon)>0$. This can be seen from an analysis of convergence of the corresponding integrals over Feynman parameters. Thus, there is no domain in the complex plane of $\varepsilon$ where both terms are given by convergent integrals. In fact, using auxiliary subtraction operators, it is possible to write down the result of expansion in such a way that both terms on the right-hand side will be convergent in some domain of $\varepsilon$. However. I prefer to follow the prescription which is implied in practice: to evaluate every term in the result for expansion in a domain of $\varepsilon$ where it is convergent and then analytically continue the corresponding result to some desirable domain.
 
Expansion by regions has the status of experimental mathematics. Usually, when studying a given limit, one starts from one-loop examples, checks results by independent methods and, finally, one understands which regions are relevant to the limit and that one obtains reliable expansion within this strategy.
Beneke provided a one-parametric example showing explicitly how expansion by regions works. The example
was used in Chapter~3 of~\cite{Smirnov:2002pj}. Guided by this example,
Jantzen \cite{Jantzen:2011nz} provided detailed explanations of how this strategy works 
in several two-loop examples by
starting from regions determined by some inequalities and covering the whole integration space
of the loop momenta, then expanding the integrand and then extending integration and
analyzing all the pieces which are obtained, with the hope that `readers would be convinced that 
the expansion by regions is a well-founded method'.

However, there is an important 
class of limits for which there is a mathematical proof. These are limits typical of Euclidean space:
for example, the off-shell large momentum limit and the large mass limit. In \cite{Smirnov:1990rz} 
(see also Appendix~B of~\cite{Smirnov:2002pj}) that the remainder of such expansion constructed with the help
of an operator which has the structure of the $R$-operation (i.e. renormalization at the diagrammatical level)
has the desirable order with respect to the parameter of expansion. This proof was for a general $h$-loop graph. It was similar to proofs of results on the $R$-operation and
was based on sector decompositions and a resolution of singularities in parametric integrals, with power counting
of sector variables. 
 
For this class of limit, the expansion of a given Feynman integral corresponding to a graph $\Gamma$ is given 
\cite{Chetyrkin:1988zz,Gorishnii:1989dd,Smirnov:1990rz} (see also \cite{Smirnov:1994tg} and Chapter~9 of \cite{Smirnov:2012gma})  by the following simple
formula:
\begin{equation}
G_{\Gamma} \sim \sum_{\gamma} G_{\Gamma/\gamma}\circ {\cal T}_{q_\gamma,m_\gamma} G_\gamma\;.
\end{equation}
which is written for the off-shell large-momentum limit, i.e. where a momentum $Q$ is considered
large and momenta $q_i$ as well as the masses $m_j$ are small. The sum runs over subgraphs $\gamma$ of $\Gamma$
which can be called asymptotically irreducible (AI): they are one-particle irreducible after identifying the two external vertices associated with the large external momentum $Q$. Moreover,
${\cal T}$ is the operator of Taylor expansion in internal masses and external momenta of a subgraph $\gamma$,
the symbol $\circ$ means the insertion of the polynomial obtained after this Taylor expansion into
the vertex of the reduced graph $\Gamma/\gamma$ to which $\gamma$ is reduced.

In the case of limits typical of Euclidean space, there is a natural one-to-one correspondence between AI subgraphs and regions in the description of the expansion within expansion by regions, so that we obtain an indirect justification 
of expansion by regions for such limits.
The set of relevant regions exactly corresponds to the set of AI subgraphs.
There are two kind of regions for each loop momentum: small and large. For a given AI subgraph $\gamma$, the corresponding
region is defined by considering each loop momentum of $\gamma$ as large and the rest of the loop momenta of $\Gamma$
(i.e. loop momenta of $\Gamma/\gamma$) as small.
For example, two subgraphs are AI for Fig.~\ref{fig:1}: the graph $\Gamma$ and the subgraph consisting of the massless line. As a result, we obtain the same contributions as above.

For limits typical of Minkowski space, to reveal the set of relevant regions is not so simple.
For example, for the threshold limit in the case where the threshold in the $q$ channel is at 
$q^2=4m^2$ and the small expansion parameter is introduced by $y=m^2-q^2/4\to 0$, the following four kind of regions for a loop momentum are relevant \cite{Beneke:1997zp}:
\begin{eqnarray}
\mbox{(hard),} && k_0\sim \sqrt{q^2}\,,\,\,\vec{k}\sim \sqrt{q^2}\,, \nonumber \\
\mbox{(soft),} && k_0\sim \sqrt{y}\,,\,\,\vec{k}\sim\sqrt{y}\,,\nonumber\\
\mbox{(potential),} && k_0\sim y/\sqrt{q^2}\,,\,\,\vec{k}\sim\sqrt{y}\,,
\nonumber \\
\mbox{(ultrasoft),} &&  k_0\sim y/\sqrt{q^2}\,,\,\,\vec{k}\sim y/\sqrt{q^2}\,. \nonumber
\end{eqnarray}
where $q=(q_0,\vec{0})$.

An alternative version of expansion by regions was formulated and illustrated via examples in
\cite{Smirnov:1999bza} within the well-known Feynman parametric representation.
This representation in the case of propagators with $-k^2$ propagators with general indices $a_i$ (powers of
the propagators) is
\begin{eqnarray}
G(q_1,\ldots,q_n;d) & =&
\left(i\pi^{d/2} \right)^h
\frac{\Gamma(\sum a-h d/2)}{\prod_i \Gamma(a_i)}
\nonumber \\ && \hspace*{-30mm}
\times
\int_0^\infty \ldots\int_0^\infty \,
\delta\left( \sum x_i-1\right) \, \prod x_i^{a_i-1} \,
U^{a-(h+1) d/2}
F^{h d/2-a} {\rm d} x_1 \ldots {\rm d} x_n
\label{Fp}
\end{eqnarray}
where $n$ is the number of lines (edges), $a=\sum a_i$, $h$ is the number of loops of the graph,
\begin{equation}
F= -V +U\sum m^2_l x_l \;,
\end{equation}
and $U$ and $V$ are two basic functions 
(Symanzik polynomials, or graph polynomials) for the given graph,
\begin{eqnarray}
U& = & \sum_{T\in T^1} \prod_{l{\not\!\, \in}  T} x_l
 \;,
\label{Dform}  \\
V&=& \sum_{T\in T^2}  
\prod_{l{\not\!\, \in}  T} x_l
\left( q^T\right)^2
 \; .
\label{Aform}
\end{eqnarray}
In (\ref{Dform}), the sum runs over
trees
of the given graph,
and, in (\ref{Aform}),
over {\em 2-trees}, i.e. subgraphs that do not involve loops
and consist of two connectivity components; $\pm q^T$ is
the  sum of the external momenta that flow
into one of the connectivity components of the 2-tree $T$. 
The products of the Feynman parameters involved are taken
over the lines that do not belong to a given tree or a 2-tree $T$.
As is well known, one can choose the sum in the argument of the delta-function
over any subset of lines. In particular, one can choose just one Feynman parameter,
$x_l$, and then the integration will be over the other parameters at $x_l=1$.
The functions $U$ and $V$ are homogeneous with respect to Feynman parameters, 
with the homogeneity degrees $h$ and
$h+1$, respectively.
 
One can consider quite general limits for a Feynman integral which depends on external momenta $q_i$
and masses and is a scalar function of kinematic invariants and squares of masses, $s_i$,
and assume that each  $s_i$ has certain scaling
$\rho^{\kappa_i}$ where $\rho$ is a small parameter. 
 
An algorithmic way to reveal regions relevant to a given limit was found in \cite{Pak:2010pt}.
It is based on the geometry of polytopes connected with the basic functions $U$ and $F$ in (\ref{Fp}).
This was a real breakthrough, both in theoretical and practical sense because, on the one hand, it became possible to
formulate expansion by regions in an unambiguous mathematical language and, on the other hand,
the authors of \cite{Pak:2010pt} presented also a public code {\tt asy.m} which was later successfully applied in various problems with Feynman integrals. 

Ironically, this algorithm and the code didn't find, in this first version,
the potential region for the threshold expansion. Later, this algorithm was updated 
and, in its current version,
it can reveal potential region as well as Glauber region. This was done by introducing an additional
decomposition of the integration domain and introducing new variables. Consider, for example,
one-loop diagram with two massive lines in the threshold limit $y=m^2-q^2/4\to 0$
\begin{equation}
  G(q^2,y)=i\pi^{d/2} \, \Gamma(\varepsilon)
  \int_0^\infty \int_0^\infty \frac{(x_1+x_2)^{2\varepsilon-2}\;\delta\left(x_1+x_2-1\right)\;
  {\rm d} x_1{\rm d} x_2
  }{\left[
  \frac{q^2}{4}(x_1-x_2)^2 + y(x_1+x_2)^2
  -i 0\right]^{\varepsilon}} \;.
  \label{prop1_alpha}  
\end{equation}

The code {\tt asy.m} in its first version revealed only the contribution of the hard region,
i.e. $x_i\sim y^0$. To make the potential region visible, let us 
decompose integration over $x_1\leq x_2$ and  $x_2\leq x_1$, with equal contributions.
In the first domain, let us turn to new variables by $x_1=x_1'/2,\;x_2=x'_2+x'_1/2$
and arrive at
\begin{equation}
  i\pi^{d/2} \, \frac{\Gamma(\varepsilon)}{2}
   \int_0^\infty \int_0^\infty \frac{(x_1+x_2)^{2\varepsilon-2}\;\delta\left(x_1+x_2-1\right)\;
   {\rm d} x_1{\rm d} x_2}
   {\left[
   \frac{q^2}{4}x_2^2 + y(x_1+x_2)^2
   -i 0\right]^{\varepsilon}}\;.
   \label{prop1_alpha_1} \nonumber
 \end{equation}
 
Now we observe two regions with the scalings  $(0, 0)$ and $(0, 1/2)$.
The second one, with $x_1\sim y^0, x_2\sim \sqrt{y}$, gives
\begin{equation}
  i\pi^{d/2} \, \frac{\Gamma(\varepsilon)}{2}
   \int_0^\infty
   \frac{{\rm d} x_2}{\left(\frac{q^2}{4}x_2^2 + y\right)^{\varepsilon}}
 =  i\pi^{d/2}\frac{1}{2} \Gamma(\varepsilon-1/2)
   \sqrt{\frac{\pi y}{q^2}} y^{-\varepsilon}
   \;.
   \label{prop1_alpha_p0} \nonumber
 \end{equation}
Taking into account that we have two identical contributions after the above decomposition,
we obtain a result for the potential contribution equal to the previous expression with omitted
$1/2$.

Observe that the expression for the function $F$ in the Feynman parametric representation 
is non-negatively
defined and only some individual terms are negative but this brings problems when looking for potential contributions.
In the current version of the code {\tt asy.m}, one can get rid of the negative terms due to additional decompositions and introduction of new variable.
Let me emphasize that this code can work successfully also in situations with a function $F$ not positively defined even without
additional decompositions -- see, e.g. \cite{Henn:2014lfa,Caola:2014lpa}.

For completeness, let me refer to \cite{Ananthanarayan:2018tog,Mishima:2018olh} where two specific ways of dealing with expansion by region
were applied.
 
Let us realize that the very word `region' is used within the strategy under discussion in a physical rather mathematical way.
By region, we mean some scaling behaviour of parameters involved.
I will present expansion by regions in a mathematical language in 
the next section using
another form of parametric representation, rather than (\ref{Fp}) and illustrate it through simple examples.

\section{Geometrical formulation}
\label{sec:2}
 
Lee and Pomeransky \cite{Lee:2013hzt} have recently derived another form of parametric representation which turns out to be
preferable in certain situations
\begin{eqnarray}
G(q_1,\ldots,q_n;d) & =&
\left(i\pi^{d/2} \right)^h
\frac{\Gamma(d/2)}{\Gamma((h+1)d/2-a) \prod_i \Gamma(a_i)}
\nonumber \\ && \hspace*{-0mm}
\times
\int_0^\infty \ldots\int_0^\infty \,  \prod_i x_i^{a_i-1} \,
P^{-\delta}  {\rm d} x_1 \ldots {\rm d} x_n\;,
\label{L-P}
\end{eqnarray}
where $\delta=2-\varepsilon$ and $P= U +F$.   
One can obtain (\ref{Fp}) from (\ref{L-P}) by \cite{Lee:2013hzt}
inserting  $1 = \int \delta(\sum_i x_i -\eta){\rm d} \eta$,  scaling $x \to \eta x$ 
and integrating over $\eta$.
  
The parametric representation takes now a very simple form: up to general powers of the integration variables, there is only
one polynomial raised to a general complex power. I believe that the fact that this function is the sum
of the two basic functions in Feynman parametric representation is not crucial and expansion by regions holds for any polynomial.
 
Let us formulate, following \cite{Semenova:2018cwy}, expansion by regions for integral (\ref{L-P})
with a polynomial with positive coefficients in the case of
limits with two kinematic invariants and/or masses of essentially different scale, where one introduces one parameter, $t$,
which is the ratio of two scales and is considered small. 
These can be such limits typical of Minkowski space as
the Regge limit, with $t\ll s$ and various versions of the Sudakov limit.
Then the polynomial in Eq.~(\ref{L-P}) is a function of Feynman parameters and $t$,
\begin{equation} 
P(x_1,\ldots,x_n,t)=\sum_{w\in S} c_{w}x_1^{w_1}\ldots x_n^{w_n} t^{w_{n+1}}\;,
\label{polynomial}
\end{equation} 
where $S$ is a finite set of points $w=(w_1,...,w_{n+1})$ and $c_{w}>0$. 

By definition, the Newton polytope $\mathcal{N}_P$ of $P$ 
is the convex hull of the points $w$ in the $n+1$-dimensional Euclidean space $R^{n+1}$
equipped with the scalar product $v\cdot w=\sum_{i=1}^{n+1} v_i w_i$. 
A facet of $P$ is a face of maximal dimension, i.e. $n$. 

{\bf The main conjecture. (Expansion by regions.)} 
The expansion of (\ref{L-P}) in the limit  
$t\to +0$ is given by  
\begin{eqnarray}
G(t,\varepsilon)\sim \sum_{\gamma}
\int_0^\infty\ldots\int_0^\infty
\left[
M_{\gamma} \left(P(x_1,\ldots,x_n,t)\right)^{-\delta}
\right]
 {\rm d} x_1 \ldots {\rm d} x_{n} \,,
\label{rhs}
\end{eqnarray}
where the sum runs over facets of the Newton polytope $\mathcal{N}_P$ of $P$, 
for which the normal vectors
$r^{\gamma} = (r^{\gamma}_1,\ldots,r_n^{\gamma},r_{n+1}^{\gamma})$, oriented inside the polytope
have $r_{n+1}^{\gamma}>0$. Let us normalize these vectors by $r_{n+1}^{\gamma}=1$. 
Let us call these facets {\em essential}.

The contribution of a given essential facet is defined by the change of variables $x_i \to  t^{r^{\gamma}_i} x_i$
in the integral (\ref{L-P}) and expanding the resulting integrand in powers of $t$.
Let us write this procedure explicitly.
For a given essential facet ${\gamma}$, the polynomial $P$ is transformed into
\begin{eqnarray}
P^{\gamma}(x_1,\ldots,x_n,t)=P(t^{r^{\gamma}_{1}} x_1,\ldots,t^{r^{\gamma}_{n}} x_n,t)
\equiv 
\sum_{w\in S} c_{w} x_1^{w_1}\ldots x_n^{w_n} t^{w\cdot r^{\gamma}}   \;.
\label{Pgm}
\end{eqnarray}
The scalar product $w\cdot r^{\gamma}$ is proportional to the projection of the point  $w$ on the vector $r^{\gamma}$.
For $w\in S$, it takes a minimal value for all the points belonging to the considered facet $w\in S \cap \gamma$.
Let us denote it by $L(\gamma)$.

The polynomial (\ref{Pgm}) can be represented as
\begin{eqnarray}
t^{L(\gamma)}\left(
P^{\gamma}_0(x_1,\ldots,x_n) + P^{\gamma}_1(x_1,\ldots,x_n,t)\right)
\label{Pgm1}\;,
\end{eqnarray}
where
\begin{eqnarray}
P^{\gamma}_0(x_1,\ldots,x_n) &=&\sum_{w\in S\cap \gamma} c_{w} x_1^{w_1}\ldots x_n^{w_n}\,,
\label{P1}
\\
P^{\gamma}_1(x_1,\ldots,x_n,t)&=&\sum_{w\in S\setminus \gamma} 
c_{w} x_1^{w_1}\ldots x_n^{w_n} t^{w\cdot r^{\gamma}-L(\gamma)} 
\label{P2}\;.
\end{eqnarray}
The polynomial $P^{\gamma}_0$ is independent of $t$ while $P^{\gamma}_1$ can be represented as a linear combination 
of positive rational powers of $t$ with coefficients which are polynomials of $x$.

For a given facet $\gamma$, let us define the operator  
\begin{eqnarray}
M_{\gamma} \left(P(x_1,\ldots,x_n,t)\right)^{-\delta}
&=& t^{\sum_{i=1}^n r^{\gamma}_i-L(\gamma)\delta} {\cal T}_{t}
\left( P^{\gamma}_0(x_1,\ldots,x_n) +  P^{\gamma}_1(x_1,\ldots,x_n,t) 
\right)^{-\delta}
\nonumber \\  \hspace*{-82mm}
&=& t^{\sum_{i=1}^n r^{\gamma}_i-L(\gamma)\delta}\ \left( P^{\gamma}_0(x_1,\ldots,x_n)\right)^{-\delta}+\ldots 
\label{M-operator}
\end{eqnarray}
where ${\cal T}_{t}$ 
performs an expansion in powers of $t$ at  $t= 0$.

{\bf Comments.}  
\begin{itemize}
\item 
An operator $M_{\gamma}$ can equivalently be defined by introducing a parameter $\rho_{\gamma}$,
replacing $x_i$ by $\rho^{r^{\gamma}_{i}} x_i$ , pulling an overall power of $\rho_{\gamma}$,
expanding in $\rho_{\gamma}$ and setting $\rho_{\gamma}=1$ in the end. 
\item 
The leading order term of a given facet $\gamma$ corresponds to the leading order of the operator
$M^0_{\gamma} $:
\begin{eqnarray}
&& \int_0^\infty\ldots\int_0^\infty
\left[
M^0_{\gamma} \left(P(x_1,\ldots,x_n,t)\right)^{-\delta}
\right]
 {\rm d} x_1 \ldots {\rm d} x_{n} 
 \nonumber \\
&=&t^{-L(\gamma)\delta+\sum_{i=1}^n r^{\gamma}_i}\, \int_0^\infty \ldots\int_0^\infty \,
\left(P^{\gamma}_0(x_1,\ldots,x_n)\right)^{-\delta}  {\rm d} x_1 \ldots {\rm d} x_n\;.
\label{L-P-LO}
\end{eqnarray}
\item
In fact, with the above definitions, we can write down the equation of the hyperplane generated by a given facet $\gamma$
as follows
\begin{eqnarray}
w_{n+1}=-\sum_{i=1}^n r^{\gamma}_i w_i+L(\gamma)\;. 
\label{hyperplane_gamma}
\end{eqnarray}
\item
Let us agree that the action of an operator $M_\gamma$ on an integral reduces to the action of
$M_\gamma$ on the integrand described above.
Then we can write down the expansion in a shorter way,
\begin{eqnarray}
G(t,\varepsilon)\sim \sum_{\gamma} M_{\gamma}  G(t,\varepsilon)
\label{rhs-ec}
 \end{eqnarray}
\item
In the usual Feynman parametrization (\ref{Fp}),
the expansion by regions in terms of operators $M_{\gamma}$ is formulated in a similar way, and this is exactly
how it is implemented in the code {\tt asy.m}~\cite{Pak:2010pt}. The expansion can be written in the same form
(\ref{rhs-ec}) but the operators $M_{\gamma}$ act on the product of the two basic polynomials 
$U$ and $F$ raised to certain powers present in (\ref{Fp}). Now, each of the two polynomials is decomposed
in the form (\ref{Pgm1}) and so on.
\item
Of course, prescriptions based on representation (\ref{L-P}) are algorithmically preferable because
the degree of the sum of the two basic polynomials is smaller than the degree of their product
$U F$ (previously used in {\tt asy.m}) so that looking for facets of the corresponding Newton polytope
becomes a simpler procedure\footnote{In fact, this step is performed within {\tt asy.m} with the help
of another code {\tt qhull}. It is most time-consuming and can become problematic in higher-loop
calculations.}. Therefore, the current version of the code {\tt asy.m} included in {\tt FIESTA} \cite{Smirnov:2015mct} (called with the command {\tt SDExpandAsy} 
is now based on this more effective procedure.
\item
It is well known that dimensional regularization might be insufficient to regularize individual contributions
to the asymptotic expansion. As it was explained in the discussion after equation~(\ref{exp2}), the natural way to overcome this problem is to introduce an auxiliary
analytic regularization, i.e. to introduce additional exponents $\lambda_i$ to powers of the propagators.
This possibility exists in the code {\tt asy.m}~\cite{Pak:2010pt}
included in {\tt FIESTA} \cite{Smirnov:2015mct}. One can choose these additional parameters in some way
and obtain a result in terms of an expansion in $\lambda_i$ followed by an expansion in $\varepsilon$.
If an initial integral can be well defined as a function of $\varepsilon$ then
the cancellation of poles in $\lambda_i$ in the sum of contributions of different regions serves as a good check of the calculational procedure, so that
in the end one obtains a result in terms of a Laurent expansion in $\varepsilon$ up to a desired order.
\end{itemize} 

\begin{figure}[h]
\includegraphics[scale=1.0]{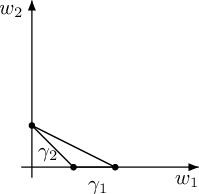}
\caption{The Newton polytope for Fig.~\ref{fig:1}.}
\label{fig:2} 
\end{figure}

To illustrate the above prescriptions let us consider a very simple example of the integral
\begin{equation}
G(t,\varepsilon)=\int_0^\infty  (x^2+x+t)^{\varepsilon-1}  {\rm d} x
\label{1parint}
\end{equation}
in the limit $t\to 0$. The polynomial involved is
$P(x,t)=\sum_{(w_1,w_2)\in S} c_{(w_1,w_2)}x^{w_1} t^{w_2}$.
The corresponding Newton polytope (triangle) is shown in Fig.~\ref{fig:2}
 
There are two essential facets $\gamma_1$ and $\gamma_2$ with the corresponding normal vectors $r_1=(0,1)$ and $r_2=(1,1)$.
For the facet $\gamma_1$, we obtain the contribution given by expanding the integrand in $t$. In the leading order, we have
\begin{eqnarray}
\int_0^\infty  (x^2+x)^{\varepsilon-1}  {\rm d} x= \frac{\Gamma (1-2 \varepsilon) \Gamma (\varepsilon)}{\Gamma (1-\varepsilon)} \;.
\end{eqnarray}
For the facet $\gamma_2,$ we obtain $t$ times the integral of the integrand with $x\to t x$ expanded in powers of $t$. 
In the leading order, we have
\begin{eqnarray}
t^{\varepsilon}\int_0^\infty  (x+1)^{\varepsilon-1}  {\rm d} x= -\frac{t^{\varepsilon}}{\varepsilon}\;.
\end{eqnarray}
The sum of the two contributions in the leading order gives
\begin{equation}
G(t,\varepsilon)\sim -\log t +O(\varepsilon) \;.
\end{equation}

\begin{figure}[htb]
\sidecaption[ht]
\includegraphics[scale=1.4]{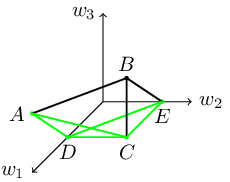}
\caption{The Newton polytope for Fig.~\ref{fig:1}.}
\label{fig:3}       
\end{figure}

Let us now consider again the example of Fig.~\ref{fig:1}. The two basic functions of Feynman parameters are
\begin{equation}
F=x_1(t(x_1+x_2)+x_2)\;,\;\;\; U=x_1+x_2\;.
\end{equation}
The set $S$ involved in the definition (\ref{polynomial}) consists of the vertices 
\[
A(2,0,1), B(1,1,1), C(1,1,0), D(1,0,0), E(0,1,0)
\]
of the Newton polytope 
for the polynomial $P= U +F$, as it is shown in Fig.~\ref{fig:3}.
   
There are two essential facets. The first one is
$CDE$ which belongs to the plane $w_3=0$ and has the normal vector $(0,0,1)$. It gives the contribution obtained by
expanding the integrand in $t$.
 
The second essential facet is
$ACD$ which belongs to the plane $w_1-w_3=1$ and has the normal vector $(-1,0,1)$.
It gives $t^{-\varepsilon}$ times the integral
\[
\frac{\Gamma(2-\varepsilon)}{\Gamma(1-2\varepsilon)}
\int_0^\infty \int_0^\infty 
x_1\left[x_1+x_1^2+x_1x_2+ tx_2+ tx_1x_2\right]^{\varepsilon-2}\, 
{\rm d}x_1 {\rm d}x_2 
\]
with the integrand expanded in $t$. Taking the leading orders in both contributions we reproduce
(\ref{f3}).

\section{Conclusion}
 
As it was argued in \cite{Semenova:2018cwy}, the more general parametric representation 
(\ref{L-P}), with a general polynomial not necessarily related to Feynman integrals, looks mathematically more natural for the proof of expansion by regions. Moreover, first steps of analysis of convergence of
integrals (\ref{L-P}) were made and expansion by regions was proven in a partial case in the leading order of expansion. Hopefully, 
expansion by regions will be sooner or later mathematically justified in the case of a general polynomial $P$.
 
Practically, expansion by regions is a very important strategy which is successfully applied for several purposes. Let me, finally, point out that one can use expansion by regions in various ways.
\begin{itemize}
\item 
One can apply the code {\tt asy.m} included in {\tt FIESTA} 
\cite{Smirnov:2015mct} (i.e. the command {\tt SDExpandAsy}) to obtain an expansion in some limit treating all the involved parameters
numerically. In particular, one can check  analytic results.
\item
One can use {\tt SDExpandAsy} with the option {\tt OnlyPrepareRegions = True} in order
to reveal relevant regions and to construct contributions to the expansion as parametric integrals which can then analytically be evaluated. Here the method of Mellin-Barnes representation can serve as an appropriate additional technique.
\item
One can study expansion in multiscale limits, applying {\tt asy.m} several times, in various orders.
\end{itemize}
 
\begin{acknowledgement}
I would like to thank the organizers of the Paris Winter Workshop ``The Infrared in QFT''
(2--6 March 2020, Paris) and the workshop ``Antidifferentiation and the Calculation of Feynman Amplitudes'' (DESY Zeuthen,4--9 October 2020) for the possibility to present this talk.
\end{acknowledgement}

%
%
%

\end{document}